# Geometry-induced electron doping in periodic semiconductor nanostructures

A. Tavkhelidze

*Ilia State University, Cholokashvili Ave. 3/5, 0162 Tbilisi, Georgia*
*E-mail address: avtotav@gmail.com*

Recently, new quantum features have been observed and studied in the area of nanostructured layers. Nanograting on the surface of the thin layer imposes additional boundary conditions on the electron wave function and induces G-doping or geometry doping. G-doping is equivalent to donor doping from the point of view of the increase in electron concentration $n$. However, there are no ionized impurities. This preserves charge carrier scattering to the intrinsic semiconductor level and increases carrier mobility with respect to the donor-doped layer. G-doping involves electron confinement to the nanograting layer. Here, we investigate the system of multiple nanograting layers forming a series of hetero- or homojunctions. The system includes main and barrier layers. In the case of heterojunctions, both types of layers were G-doped. In the case of homojunctions, main layers were G-doped and barrier layers were donor-doped. In such systems, the dependence of $n$ on layer geometry and material parameters was analysed. Si and GaAs homojunctions and GaAs/AlGaAs, Si/SiGe, GaInP/AlGAs, and InP/InAlAs heterojunctions were studied. G-doping levels of $10^{18}$-$10^{19}$ cm$^{-3}$ were obtained in homojunctions and type II heterojunctions. High G-doping levels were attained only when the difference between band gap values was low.

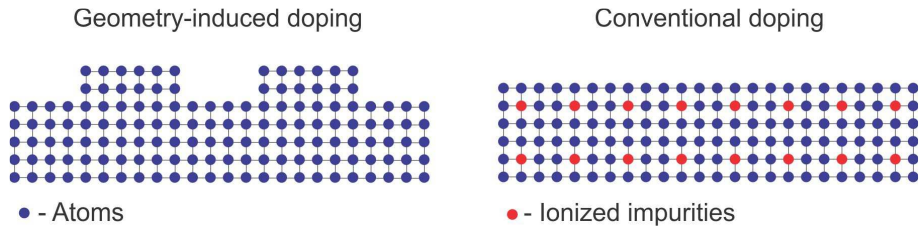

Geometry-induced doping     Conventional doping

● - Atoms     ●● - Ionized impurities

## 1. Introduction

Developments in nanotechnology allow the fabrication of densely packed, periodic structures [1-4]. Recently, ultra short period nanopore arrays and nanogratings have been obtained by block copolymer lithography [1, 2]. Another method for nanograting fabrication is a multi-beam interference lithography. Using these techniques, gratings with 10-nm [3] and even sub-10 nm pitch [4] have been fabricated. At the same time, nanograting (NG) has been shown to dramatically improve thermoelectric [5] and electron emission properties [6] when the grating pitch becomes comparable with the electron's de Broglie wavelength. This is due to the special boundary conditions imposed by NG on the electron wave function. Supplementary boundary conditions forbid some quantum states, and the density of quantum states (DOS) is reduced (in all bands). Electrons rejected from NG-forbidden quantum states have to occupy empty states with a higher energy $E$. Fermi energy $E_F$ increases, and the electronic properties of the NG layer change. In the case of semiconductor materials, electrons rejected from the valence band (VB) occupy empty quantum states in the conduction band (CB). Electron concentration $n$ in the CB increases, which can be termed as geometry-induced electron doping or G-doping. G-doping is equivalent to donor doping from the point of view of the increase in $n$ and Fermi energy $E_F$. However, there are no ionized impurities. This maintains charge carrier scattering to the intrinsic semiconductor level and increases carrier mobility with respect to the donor-doped layer of the same electron concentration. G-doping is temperature independent because it originates from layer geometry and no ionized impurities are involved.

Other methods of doping without impurities include the well-known modulation doping and the recently introduced polarization doping [7]. Both are 2D in nature. However, a 3D approach to modulation doping was introduced in [8] to improve the thermoelelectric characteristics of nanocomposites [9, 10]. The influence of periodic structures on electronic properties has been studied in related geometries, such as periodic curved surfaces [11, 12], nanotubes [13], cylindrical surfaces with nonconstant diameter [14], and strain-driven nanostructures [15].

Electron confinement to the NG layer is needed to obtain G-doping. The layer can be made freestanding, but in practice it is usually sandwiched between wide bandgap layers. The NG layer thickness is fundamentally limited by the requirement of having quantum properties. However, thin layers have low optical absorption. The layer thickness also limits the lateral charge and heat transport.

Here, we investigate G-doping in multiple nanograting layers. Such layers are quasi-3D and have improved optoelectronic [16] and thermoelectric characteristics with respect to a single layer. Multiple nanograting layers consist of a replicated structure including main and barrier layers, forming a series of hetero- or homojunctions. The main layer is thicker than the barrier layer and plays a leading role in carrier transport and optical absorption. The barrier layer forms electron confinement energy regions. It also contributes to carrier transport and optical absorption. Both the main and barrier layers have NG geometry.

The objectives of this work are to calculate $n$ and $E_F$ in multiple nanograting layers and to find out how they are dependent on layer thicknesses and material properties. First, we introduce G-doping in a single nanograting layer (Sec. 2). Next, we calculate $n$ and $E_F$ in a homojunction multilayer structure (Sec. 3). Subsequently, we calculate the same for a heterojunction multilayer structure (Sec. 4). Finally, the possibility of realizing such structures and their advantages for optoelectronic and thermoelectric devices are discussed (Subsection 4.4). In Sec. 5, our conclusions are summarized briefly. An analysis is made within the limits of a parabolic band, a wide quantum well, and degenerate electron gas approximations. A parabolic band approximation can be used as we consider only band edges. Wide quantum well is a good approximation as we regard relatively thick layers with better optical absorption and lateral transport properties. Degenerate electron gas approximation is suitable as we consider only high electron concentrations (high G-doping levels).

## 2. DOS and electron concentration in a single nanograting layer

Figure 1 shows a cross section of a single NG layer.

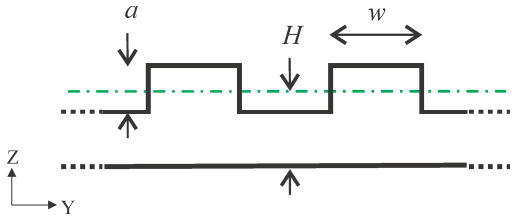

Fig. 1  Cross section of nanograting layer.

The grating has depth $a$ and period $2w$. To make a comparison, we set the reference layer as a plain layer with thickness $H$ such that it has the same cross-section area. Nanograting imposes additional boundary conditions on the electron wave function and forbids some quantum states. The DOS in energy $\rho(E)$ is reduced [17] with respect to the reference well

$$\rho(E) = \rho_0(E)/G \,, \qquad (1)$$

where $\rho_0(E)$ is the DOS (number of quantum states within the unit energy region and in the unit volume) in a reference well, and $G = G(H, w, a) > 1$ is the geometry factor.

The geometry factor or DOS calculation requires solving the time-independent Schrödinger equation in NG geometry. Mathematically, there is no difference between DOS reduction and electromagnetic (TM) mode depression [18, 19]. The Helmholtz equation and Dirichlet boundary conditions are used in both cases. Unfortunately, there is no exact analytical solution for NG geometry. The approximate analytical expression known as Weyl's formula [20, 21] allows the calculation of DOS by using a ratio of layer surface area and volume. The perturbation method [22] has been used to find an approximate analytical expression for $G$ in NG geometry. The DOS for electromagnetic modes in resembling geometries has been numerically calculated in the literature related to the Casimir effect [23].

Nanograting reduces the DOS in all bands. Electrons rejected from NG-forbidden quantum states have to occupy empty quantum states with higher E. In a semiconductor, electrons rejected from the VB have to occupy empty (and not forbidden by NG) energy levels in the CB (Fig. 2). Electrons are rejected from low energy levels and occupy high energy ones. During this process, the Fermi energy increases from $E_F^{(0)}$ to $E_F$. To simplify the presentation, in Fig. 2, we presume that $T=0$ ($T$ is the absolute temperature) and the energy levels are equidistant on the energy scale (the geometry-induced energy level shift is also ignored). We use Eq. (1) to calculate the DOS and investigate the $G$-dependence of $n$ and $E_F$. The density of the NG-forbidden quantum states is

$$\rho_F(E) = \rho_0(E) - \rho_0(E)/G = \rho_0(E)(1 - 1/G) \,. \qquad (2)$$

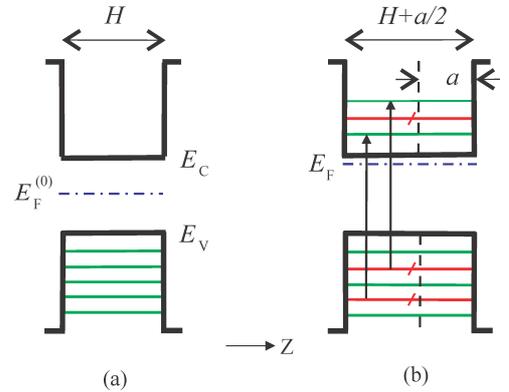

Fig. 2  Energy diagrams of a) the reference semiconductor quantum well and b) the NG layer. The horizontal (green) lines indicate occupied energy levels; the crossed (red) lines indicate NG-forbidden energy levels.

To determine the number of rejected electrons $n_r$ (per unit volume), Eq. (2) should be integrated over the electron confinement energy region.

$$n_r = \int_{con} dE \, \rho_F(E) = (1 - 1/G) \int_{con} dE \, \rho_0(E) \,. \qquad (3)$$

Here, we assume that electron confinement takes place only in narrow energy intervals inside which $G$ is energy-independent. In most cases, the thin layer is grown on a semiconductor substrate, and the confinement energy regions are band offsets (Fig 3).

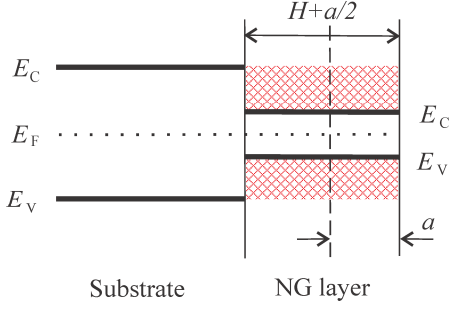

Fig. 3 Electron confinement energy ranges (red) in the system of the substrate and NG layer.

To obtain electron confinement in multiple NG layers, transitional barrier layers are required. We consider cases of wide bandgap material forming a heterojunction and the same material forming a homojunction. In the case of the heterojunction, the confinement regions are band discontinuities (both type-I and type-II band alignments are considered). In the case of the homojunction, the confinement regions originate from barrier layer donor doping.

## 3. Electronic properties of multiple homojunction NG layer

We begin by calculating $n$ in a homojunction NG layer, which is relatively straightforward. A homojunction is formed between the intrinsic main layer and a donor-doped barrier layer. Let us consider a case in which all layers have a plain geometry (NG is not yet formed). Figure 4a shows an energy diagram of such structure (typically obtained by homoepitaxal growth) [24]. Charge depletion regions form inside both the i-type main layer and the n-type barrier layer. The band edges curve, shaping the confinement regions in the main layer VB and in the barrier layer CB. Electrons with energies $E_V - \Delta E_{con}^{(0)} < E < E_V$ are confined to the main layers, and those with energies $E_C - \Delta E_{con}^{(0)} < E < E_C$ are confined to the barrier layers. Here, we assume that both the main and barrier layers are relatively thick such that the electron wave functions do not overlap and we can ignore the mini-band formation.

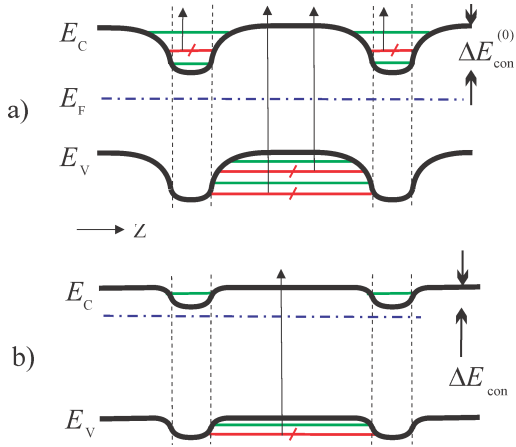

Fig. 4 Homojunction structure energy diagrams: a) before NG formation, b) after NG formation. The horizontal (green) lines indicate occupied energy levels; the crossed (red) lines indicate NG-forbidden energy levels.

### 3.1 Influence of Nanograting

Next, we fabricate NG on the surface of all layers simultaneously (the fabrication of such structure is discussed later in this work). NG forbids some energy levels (red lines) and induces G-doping depending on the geometry factor value. We introduce the two geometry factors $G_m$ and $G_b$ for the main and barrier layers, respectively (the layers have different geometry factors owing to their varying thicknesses).

If $G_m > 1$ some electrons are rejected from the VB of the main layer (Fig. 4). If $G_b > 1$ some electrons are rejected from the CB of the barrier layer and occupy higher energy levels in the same band. First, let us consider the cases of $G_m > 1$ and $G_b = 1$. When $G_m > 1$, the Fermi level increases in the main layer, which is equivalent to shifting band edges down on the energy scale. When $G_b = 1$, the band edges remain intact in the barrier layer. With the downward shift in the main layer, the VB confinement region will shrink, and fewer electrons will be rejected from the CB. This will continue until the rejected electrons achieve some equilibrium value. In other words, the downward shift of the band edge provides negative feedback to the G-doping process. Next, let us consider the cases of $G_m > 1$ and $G_b > 1$. Here, the barrier layer band edges also shift downward, which leads the confinement energy regions to widen. This increases the number of rejected electrons in the main layer. In other words, barrier layer G-doping provides positive feedback to main layer G-doping. Finally, we have a combination of negative feedback originating from the main layer and positive feedback originating from the barrier layer. In the case of overall negative feedback, the G-doping of the two layers will become saturated, and an equilibrium value of $\Delta E_{con}$ will be attained (Fig. 4b). The equilibrium value depends on $G_m$, $G_b$, and semiconductor material parameters.

### 3.2 Electron confinement energy region

To find $\Delta E_{con}$ we first have to calculate the electron concentration in both the main and barrier layers. Applying Eq. (2) to the main layer, we find the number of rejected electrons

$$n_r = (1 - 1/G_m) \int_{E_V}^{E_V - \Delta E_{con}} dE\, \rho_{0V}(E), \qquad (4)$$

Where $\rho_{0V}(E)$ is the initial (to NG) DOS in the main layer VB. Let us consider the case in which the number of rejected electrons is much higher than the initial number of electrons in the CB, $n_r \gg n_0$. Then, the CB electron concentration will be merely $n_r$. Further, we use the following well-known expression of the valence band DOS to calculate $n_r$ [25] :

$$\rho_{0V}(E) = 2\pi^{-1/2} N_V (k_B T)^{-3/2} (E - E_V)^{1/2}. \qquad (5)$$

Here, $N_V$ is the effective VB density of states. Inserting Eg. (5) in Eq. (4) and integrating (zero on an energy scale at $E_V = 0$), we find

$$n_r = (4/3\sqrt{\pi}) N_V (1 - 1/G_m)(\Delta E_{con}/k_b T)^{3/2}. \qquad (6)$$

Further, we determine the Fermi levels in both layers by using the corresponding electron concentrations. According to the well-known expression for degenerate semiconductors [26], the Fermi level in the main layer can be found by

$$E_F - E_C^{(m)} = k_B T \left[ \ln(n_r/N_C) + 2^{-3/2}(n_r/N_C) \right], \qquad (7)$$

where $N_C$ is the effective CB density of states. The CB electron concentration of the barrier layer does not change during NG formation because there are no electrons transferred from the VB to the CB. However, the Fermi level in the barrier layer moves up on the energy scale owing to the NG-induced reduction in $N_C$

$$E_F - E_C^{(b)} = k_B T \left[ \ln(G_b n_d/N_C) + 2^{-3/2}(G_b n_d/N_C) \right]. \qquad (8)$$

Here, $n_d$ is the initial electron concentration in the barrier layer (obtained by donor doping). The energy diagram (Fig. 4) shows that $\Delta E_{con} = E_C^{(m)} - E_C^{(b)}$. Subtracting Eq. (7) from Eq. (8) and inserting the last expression, we find

$$(\Delta E_{con}/k_B T) = \ln(G_b n_d/n_r) + 2^{-3/2}[(G_b n_d - n_r)/N_C]. \qquad (9)$$

Inserting Eq. (6) in Eq. (9) and setting $\eta \equiv \Delta E_{con}/k_B T$, we get the following nonlinear equation for $\eta$ :

$$\eta + \frac{3}{2} \ln \eta + 2^{-3/2} \frac{4}{3\sqrt{\pi}} \frac{N_V}{N_C} \frac{G_m - 1}{G_m} \eta^{3/2} - \ln \frac{3\sqrt{\pi} G_m G_b n_d}{4 N_V (G_m - 1)} - 2^{-3/2} \frac{G_b n_d}{N_C} = 0 . \qquad (10)$$

Equation (10) was solved numerically to find $\Delta E_{con}$ by using two geometry factors and the $n_d$, $N_C$, and $N_V$ values for Si and GaAs materials.

### 3.3 Electron concentration and Fermi level

The obtained values of $\Delta E_{con}$ were inserted in Eq. (6) to find $n_r$. Next, we obtained the Fermi level of the main layer by inserting values of $n_r$ in Eq. (7). Formula Eq. (8) was used to find the Fermi level position in the barrier layer. Table 1 shows the results for geometry factors $G_m = 1.02$, $G_b = 1.1$, and T=300 K. For barrier donor-doping levels of $10^{18}$-$10^{20}$ cm$^{-3}$ and $G$ values rather close to unity ($G_m = 1.02$, $G_b = 1.1$), main layer G-doping levels of $10^{17}$-$10^{18}$ cm$^{-3}$ were obtained. We chose these ranges of G-doping levels because they are frequently used in applications. Higher and lower G-doping levels can be obtained as well.

Table 1. Electron concentration and Fermi levels in the main and barrier layers for Si and GaAs materials. Energy was measured from the corresponding layer CB edge.

| Material | $n_d$ [cm$^{-3}$] | $\Delta E_{con}$ [meV] | $n_r$ [cm$^{-3}$] | $E_F^{(m)}$ [meV] | $E_F^{(b)}$ [meV] |
|---|---|---|---|---|---|
| Si | $3 \times 10^{18}$ | 45 | $6 \times 10^{17}$ | 103 | 58 |
| | $1 \times 10^{19}$ | 64 | $1 \times 10^{18}$ | 89 | 25 |
| | $3 \times 10^{19}$ | 87 | $1 \times 10^{18}$ | 77 | -10 |
| | $1 \times 10^{20}$ | 126 | $2.8 \times 10^{18}$ | 62 | -64 |
| GaAs | $1 \times 10^{18}$ | 47 | $3.2 \times 10^{17}$ | 3.4 | -44 |
| | $3 \times 10^{18}$ | 86 | $8 \times 10^{17}$ | -29 | -115 |
| | $5 \times 10^{18}$ | 119 | $1.3 \times 10^{18}$ | -52 | -172 |
| | $1 \times 10^{19}$ | 197 | $2.8 \times 10^{18}$ | -110 | -297 |

Table 1 shows that one-order-higher donor doping $n_d$ ($10^{20}$ cm$^{-3}$) is needed in Si with respect to GaAs ($10^{19}$ cm$^{-3}$) to obtain a G-doping level of $2.8 \times 10^{18}$ cm$^{-3}$. It is convenient to interpret this result using terms of negative and positive feedback. The negative feedback in the main layer is weaker for Si material because it has a higher $n_0$ value compared to GaAs. Owing to this high $n_0$, the band edges shift downward less rapidly with increasing $n_r$, following the well-known logarithmic dependence $\Delta E_F = k_B T \ln(n_r/n_0)$. Because $\Delta E_F$ being positive in our case, a higher $n_0$ result in weaker negative feedback. Weaker negative feedback in the main layer requires weaker positive feedback in the barrier layer to obtain equilibrium. Positive feedback is weaker in highly donor-doped barrier layers (because shifting band edges require more rejected electrons). This explains why a higher $n_d$ is required in the Si barrier layer compared to GaAs.

As Figure 3 shows, there are potential barriers for electrons in the CB and holes in the VB of multilayer NG. These affect carrier transport in the Z direction. However, the barrier height is $\Delta E_{con}$, which is of the order of a few $k_B T$, as Table 1 indicates (except for very high levels of donor doping, $k_B T = 26$ meV for T=300 K). Charge carriers easily overcome such obstacles (thermionic emission), and their influence can be ignored. In the case of high values of $\Delta E_{con}$, barriers may block carrier transport in the Z direction. To avoid this, a high level of G-doping should be obtained not by increasing the donor doping of barrier layers but by increasing the $G_m$ and $G_b$ values.

During numerical calculations, we kept $E_F^{(m)}$ and $E_F^{(b)}$ in proximity to the corresponding CB edges to stay within the limits of the degenerate electron gas approximation (so that Eq. (7) and Eq. (8) are valid). For lower doping levels, other expressions should be used instead of Eqs. (7) and (8).

## 4. Electronic properties of multiple heterojunction NGL

Here, we calculate $n$ and $E_F$ in heterojunction layers. Figure 5a shows an energy diagram of NG layers

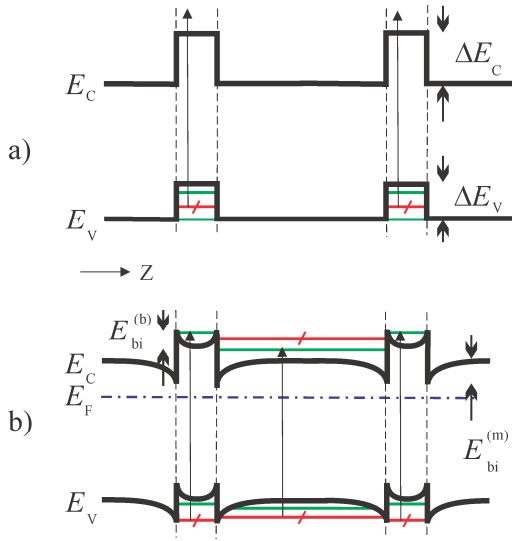

Fig. 5 Energy diagrams of heterojunction structure: a) before NG formation, b) after NG formation. The horizontal (green) lines indicate occupied energy levels; the crossed (red) lines indicated NG-forbidden energy levels

consisting of relatively thick main layers and thinner barrier layers. The band edges alignment is type II. The barrier layer bandgap is wider. The band offsets are $\Delta E_C$ and $\Delta E_V$. A narrow quantum well is formed in the barrier layer VB. We assume that both layers are thick and the sub-band formation can be ignored.

### 4.1 Influence of Nanograting

When NG is applied, some quantum states become forbidden (red lines) and G-doping is induced. First, let us consider the cases of $G_b > 1$ and $G_m = 1$. When $G_b > 1$ electrons are rejected from the barrier layer VB. The barrier layer band edges shift downward on the energy scale, and G-doping is induced. The main layer band edges remain intact when $G_m = 1$. Owing to the downward shift, the VB confinement region shrinks and fewer electrons are rejected. At the same time, a quantum well forms in the CB of the barrier layer. When $G_b > 1$, the DOS is reduced in this quantum well, and electrons rejected from the VB have to occupy higher energy levels. This further magnifies G-doping in the barrier layer. However, the barrier layer doping process will become saturated because when the VB quantum well depth drops to zero, no more electrons are rejected. Therefore, the G-doping in the barrier layer will continue until the rejected electrons reach some equilibrium value. In other words, the downward shift of the barrier layer band edges provides negative feedback. The G-doping in the barrier layer influences the main layer band edges as well. It forms n+-i contacts, and the main layer band edges curve downward near the interfaces. This curving forms a new quantum well in the VB of the main layer (Fig. 5b). Next, let us consider the cases of $G_b > 1$ and $G_m > 1$. When $G_m > 1$, electrons are rejected from the newly formed quantum well in the VB of the main layer. This induces G-doping in the main layer and leads to a downward shift of its band edges. The contract type of n+-i changes to n+-n. This reduces band edge curving near the interface in both layers. The quantum well in the barrier layer VB deepens, and the barrier layer G-doping is magnified. In other words, the main layer provides positive feedback to G-doping in the barrier layer. Finally, we have a combination of negative feedback originating from the barrier layer and positive feedback originating from the main layer. In the case of overall negative feedback, the G-doping of the two layers will become saturated, and some equilibrium values of $E_{bi}^{(m)}$ and $E_{bi}^{(b)}$ will be attained (Fig. 5b). These values depend on $G_b$, $G_m$, material parameters, and band offsets.

### 4.2 Electron confinement energy regions

To obtain the equilibrium values $E_{bi}^{(m)}$ and $E_{bi}^{(b)}$, we first have to find the number of rejected electrons in both layers. Let us use the condition of continuity of electric displacement $\varepsilon \Theta$ at the interface (Ref. 23, p.127) of an isotype heterojunction. Here, $\varepsilon$ is the dielectric constant, and $\Theta$ is the electric field. In our case, the condition of $\varepsilon \Theta$ continuity gives

$$(n_m / \varepsilon_m)\left[k_B T\left(\exp(E_{bi}^{(m)} / k_B T) - 1\right) - E_{bi}^{(m)}\right] = (n_b / \varepsilon_b) E_{bi}^{(b)}. \quad (11)$$

Here, $n_m$ and $n_b$ are the CB electron concentrations in the corresponding layers, and $\varepsilon_m$ and $\varepsilon_b$ are dielectric constants. In this expression, the exponential term accounts for the electron accumulation in the main layer near the interface. Using Eq. (6) and taking into account that the barrier layer VB quantum well depth is $\Delta E_V - E_{bi}^{(m)}$, we obtain the number of electrons rejected from the barrier layer VB as

$$n_r^{(b)} = (4/3\sqrt{\pi}) N_V^{(b)} (1 - 1/G_b)\left[\left(\Delta E_V - E_{bi}^{(b)}\right) / k_B T\right]^{3/2}. \quad (12)$$

Repeating the same procedure for the main layer and taking into account that its quantum well (VB) depth is $E_{bi}^{(m)}$, we get

$$n_r^{(m)} = (4/3\sqrt{\pi}) N_V^{(m)} (1 - 1/G_m)\left[E_{bi}^{(m)} / k_B T\right]^{3/2}. \quad (13)$$

In the cases of $n_m >> n_0^{(m)}$ and $n_b >> n_0^{(b)}$, where $n_0^{(m)}$ and

$n_0^{(b)}$ are the initial electron concentrations in the CB, we merely write

$$n_b = n_r^{(b)} \text{ and } n_m = n_r^{(m)}. \quad (14)$$

Inserting Eqs. (13) and (12) in Eq. (14) and further inserting the resulting $n_b$ and $n_m$ in Eq. (11) gives the following nonlinear equation for $E_{bi}^{(m)}$ and $E_{bi}^{(b)}$:

$$\alpha \frac{k_B T}{E_{bi}^{(b)}} \left[ \exp \frac{E_{bi}^{(m)}}{k_B T} - 1 \right] - \alpha \frac{E_{bi}^{(m)}}{E_{bi}^{(b)}} - \\ - \gamma \frac{G_m (G_b - 1)}{G_b (G_m - 1)} \left( \frac{\Delta E_V - E_{bi}^{(b)}}{E_{bi}^{(m)}} \right)^{3/2} = 0 \quad (15)$$

Here, $\alpha \equiv \varepsilon_b / \varepsilon_m$ and $\gamma \equiv N_V^{(b)} / N_V^{(m)}$.

To find the values of $E_{bi}^{(m)}$ and $E_{bi}^{(b)}$, one more equation is required. This can be obtained from the condition of equality of Fermi levels (zero external bias). Using Eq. (7) for the main and barrier layers, we have

$$E_F - E_c^{(m)} = k_B T \left[ \ln(n_r^{(m)} G_m / N_c^{(m)}) + 2^{-3/2} (n_r^{(m)} G_m / N_c^{(m)}) \right] \quad (16)$$

and

$$E_F - E_c^{(b)} = k_B T \left[ \ln(n_r^{(b)} G_b / N_c^{(b)}) + 2^{-3/2} (n_r^{(b)} G_b / N_c^{(b)}) \right]. \quad (17)$$

Subtracting Eq. (17) from Eq. (16) gives

$$\frac{E_c^{(m)} - E_c^{(b)}}{k_B T} = \ln \left[ \frac{\gamma (G_b - 1)}{\beta (G_m - 1)} \left( \frac{\Delta E_v - E_{bi}^{(b)}}{E_{bi}^{(m)}} \right)^{3/2} \right] + \\ + \frac{\sqrt{2}}{3\sqrt{\pi}} \left[ \frac{\beta \theta (G_b - 1)}{\gamma} \left( \frac{\Delta E_v - E_{bi}^{(b)}}{k_B T} \right)^{3/2} - \\ \theta (G_m - 1) \left( \frac{E_{bi}^{(m)}}{k_B T} \right)^{3/2} \right] \quad , (18)$$

where $\beta \equiv N_C^{(b)} / N_C^{(m)}$ and $\theta \equiv N_V^{(m)} / N_V^{(m)}$. At the same time, the energy diagram (Fig. 5b) shows that

$$E_C^{(m)} - E_C^{(b)} = \Delta E_C - E_{bi}^{(b)} - E_{bi}^{(m)}. \quad (19)$$

Inserting Eq. (19) in Eq. (18), we get one more nonlinear equation for $E_{bi}^{(m)}$ and $E_{bi}^{(b)}$:

$$\frac{E_{bi}^{(m)}}{k_B T} + \frac{E_{bi}^{(b)}}{k_B T} + \frac{\sqrt{2}}{3\sqrt{\pi}} \left[ \frac{\beta \theta (G_b - 1)}{\gamma} \left( \frac{\Delta E_v - E_{bi}^{(b)}}{k_B T} \right)^{3/2} - \\ \theta (G_m - 1) \left( \frac{E_{bi}^{(m)}}{k_B T} \right)^{3/2} \right] + \\ + \frac{3}{2} \ln \frac{\Delta E_v - E_{bi}^{(b)}}{E_{bi}^{(m)}} + \ln \frac{\gamma (G_b - 1)}{\beta (G_m - 1)} - \frac{\Delta E_C}{k_B T} = 0 \quad (20)$$

Equations (20) and (15) result in a system of two nonlinear equations with the two variables $E_{bi}^{(m)}$ and $E_{bi}^{(b)}$. We solved this system numerically to obtain the values of $E_{bi}^{(m)}$ and $E_{bi}^{(b)}$.

*4.3 Electron concentration and Fermi energy*

The obtained values were inserted in Eqs. (13) and (12) to find the electron concentrations $n_r^{(m)}$ and $n_r^{(b)}$. Table 2 presents the solutions of a system Eqs. (20) and (15) for type-II band alignment heterojunctions GaInP/AlGAs, InP/InAlAs, and Si/SiGe. The materials were selected such that the CB offset was not too large compared to the thermal energy at T=300 K. The material parameters for the compositions $Ga_{0.52}In_{0.48}P//Al_{0.43}Ga_{0.57}As$, $InP/In_{0.52}Al_{0.48}As$, and $Si/Si_{0.9}Ge_{0.1}$ were collected from [27-29]. The values of $G_m$ and $G_b$ were varied to obtain G-doping levels of $10^{18}$ - $10^{19} cm^{-3}$. Other material pairs were investigated as well, and it was found that a high level of G-doping can be obtained only if the difference in the bandgaps of the main and barrier layers is low. Table 2 shows that a G-doping level of $10^{18} cm^{-3}$ in the main layers, together with a G-doping level of $10^{19}$ cm$^{-3}$ in the barrier layers, can be obtained for type-II heterojunctions. Unlike in homojunctions, the doping in both the main and barrier layers is geometry-induced. However, slightly higher $G_m$ and $G_b$ values are needed to achieve such doping levels.

Table 2. Electron concentration and other parameters of NG-staggered heterojunctions.

| Materials [main/barrier] | $\Delta E_C$ [meV] | $\Delta E_V$ [meV] | $G_m$ | $G_b$ | $E_{bi}^{(m)}$ [meV] | $E_{bi}^{(b)}$ [meV] | $n_r^{(m)}$ [cm$^{-3}$] | $n_r^{(b)}$ [cm$^{-3}$] | $E_{tb}$ [meV] |
|---|---|---|---|---|---|---|---|---|---|
| $Ga_{0.52}In_{0.48}P/$ $/Al_{0.45}Ga_{0.55}As$ | 257 | 14 | 1.02 | 1.2 | 81 | 58 | 1.3 x $10^{18}$ | 1.1 x $10^{19}$ | 118 |
| $InP/$ $/In_{0.52}Al_{0.48}As$ | 34 | 25 | 1.2 | 1.8 | 131 | 111 | 1.3 x $10^{18}$ | 4.8 x $10^{19}$ | 98 |
| $Si/$ $/Si_{0.5}Ge_{0.5}$ | 56 | 24 | 1.1 | 1.006 | 28 | 15 | 1.4 x $10^{18}$ | 2.1 x $10^{18}$ | 13 |

We also calculated the type-I alignment heterojunction GaAs/Al$_x$Ga$_{1-x}$As. In this case, there was no charge accumulation layer in the CB, and the exponential term was absent in Eq. (15). Numerical calculations show that it is difficult to obtain substantial G-doping in the case of type-I alignment. The maximum electron concentration in the main layer was 1.5 x $10^{15}$ cm$^{-3}$ for the values of $x$=0.05, $G_m$=1.001, and $G_b$=1.1. We explain such result by the large difference between the bandgaps (band offsets add in type-I alignment), leading to the large difference in strength between negative and positive feedback, which complicates the achievement of equilibrium. Under these conditions, equilibrium can be reached only at very low values of band offsets, leading to low doping levels.

As Figure 4 shows, there are potential barriers for electrons in the CB and holes in the VB of multilayer NG. These barriers affect carrier transport in the Z direction. However, the barrier height in the CB is approximately $E_{tb} = \Delta E_C - E_{bi}^{(m)} - E_{bi}^{(b)}$ (Fig. 5) and corresponds the values shown in Table 2. These values do not exceed a few $k_B T$. Charge carriers easily overcome such obstacles (thermionic emission), and their influence can be ignored.

### 4.4 G-doping for high electron mobility applications

The realization of G-doping is attractive from the point of view of high mobility applications [30]. Let us use a multi-junction solar cell [31, 32] as an example to estimate the improvement of characteristics due to G-doping. In this device, the window, emitter, and tunnel junction layer doping level are roughly $10^{18}$ cm$^{-3}$. At this level of donor doping, ionized impurities reduce electron mobility by a factor of 4 in GaAs [33] (at $T$=300 K) and by a factor of 10 in Si [34]. Most types of solar cells use transparent conductive oxides [35, 36] with doping levels of $10^{20}$-$10^{21}$ cm$^{-3}$. At this donor doping level, ionized impurities reduce electron mobility by a factor of 30-50 in GaAs. Thus, using G-doping in these layers can dramatically improve the characteristics of solar cells.

Multiple NG layers can be realized by epitaxial growth on the top of lead grating, previously formed on the substrate. Such structure has been modeled [37] and fabricated [38] on GaAs substrate by using an epitaxial growth technique. Layers grown using this technique has diverse geometry compared to a single NG layer. They have gratings from both sides, and their interfaces resemble a sine instead of a square. With an increasing number of layers, the sine amplitude is reduced until plain layers are finally achieved. The geometry factor and G-doping level decrease with the increasing number of layers.

It is challenging to obtain high G values or considerable DOS reduction in both NG and similar geometries. However, as Tables 1 and 2 show, the required G-doping values are quite close to unity (especially for the main layer). Obtaining such values in multiple NG systems seems to be straightforward.

The above analysis was made with the assumption of G energy independence Eq. (3). Usually, the electron confinement energy regions are small (tens or hundreds of meV), and this assumption is valid. However, the NG layer can be made freestanding or grown on a substrate with a very wide bandgap. In these cases, the confinement energy intervals are much wider, and care should be taken when assuming G energy independence.

### 5. Conclusions

Geometry-induced electron doping (G-doping) is investigated in multiple nanograting layers. The layers are composed of main and barrier layers, forming a series of isotype homo- or heterojunctions. The barrier layers are used to form electron confinement energy regions. Both the main and barrier layers are n-type. In the case of homojunctions, the barrier layers are donor-doped to obtain electron confinement and induce G-doping in the main layers. In the case of heterojunctions, both the main and barrier layers are G-doped. Such parameters as electron concentration, Fermi level, and confinement region width are calculated for these systems. For Si and GaAs homojunctions, a main layer G-doping level of $10^{17}$ -$10^{18}$ cm$^{-3}$ is obtained at a barrier layer donor doping of $10^{18}$ -$10^{20}$ cm$^{-3}$ and geometry factor values of $G_m$=1.02 and $G_b$=1.1. One-order-higher donor doping is required in Si with respect to GaAs to obtain the same G-doping level. For high electron mobility, it is preferable to have high geometry factor values. For type-II heterojunctions Ga$_{0.52}$In$_{0.48}$P/Al$_{0.43}$Ga$_{0.57}$As, InP/In$_{0.52}$Al$_{0.48}$As, and Si/Si$_{0.9}$Ge$_{0.1}$, a main layer G-doping level of $10^{18}$ cm$^{-3}$ is obtained at a barrier layer G-doping level of $10^{18}$ -$10^{20}$ cm$^{-3}$ and geometry factor values of $G_m$=1.02-1.2 and $G_b$=1.006-1.8. It is found that a high G-doping level could be attained only when the bandgap difference is low. A G-doping level of only $10^{15}$ cm$^{-3}$ is obtained for the type-I heterojunction GaAs/AlGaAs. Such low value is explained by the addition of band offsets, resulting in large bandgap differences. Electron mobility is higher in G-doped layers compared to donor-doped layers of the same doping level. Heterojunction G-doped layers had higher electron mobility compared to homojunction ones. G-doping opens prospects for new quasi-3D optoelectronic and thermoelectric systems. At the same time, G-doping is temperature independent and can be used to extend the working temperature ranges of cryogenic and power electronics.

### Acknowledgments
The author thanks G. Japaridze for his useful discussions, and the Physics Department of New York University, where part of this work was done, for their hospitality. Work was supported by scholarship program of German Academic Exchange Service (DAAD). Apparatus received from EU TEMPUS project 530278-TEMPUS-1-2012-1-DE-TEMPUS-JPHES was used for preparation of this work.